\documentclass{article}

\usepackage{arxiv}

\usepackage[utf8]{inputenc}
\usepackage[T1]{fontenc}
\usepackage{hyperref}
\usepackage{url}
\usepackage{booktabs}
\usepackage{amsfonts}
\usepackage{amsmath}
\usepackage{nicefrac}
\usepackage{microtype}
\usepackage{xcolor}
\usepackage{graphicx}
\usepackage[authoryear,round]{natbib}
\usepackage{multirow}
\newcommand{\doi}[1]{\href{https://doi.org/#1}{\nolinkurl{doi:#1}}}
\graphicspath{ {./images/} }
\hypersetup{hidelinks}

\renewcommand{\headeright}{Applied Ocean Research 176 (2026) 105281}
\renewcommand{\undertitle}{Author version of the published article}
\renewcommand{\shorttitle}{}
\title{Multispectral representation of Distributed Acoustic Sensing data: A framework for physically interpretable feature extraction and visualisation}

\date{}

\author{
	\href{https://orcid.org/0000-0001-8883-2618}{\includegraphics[height=1.6em]{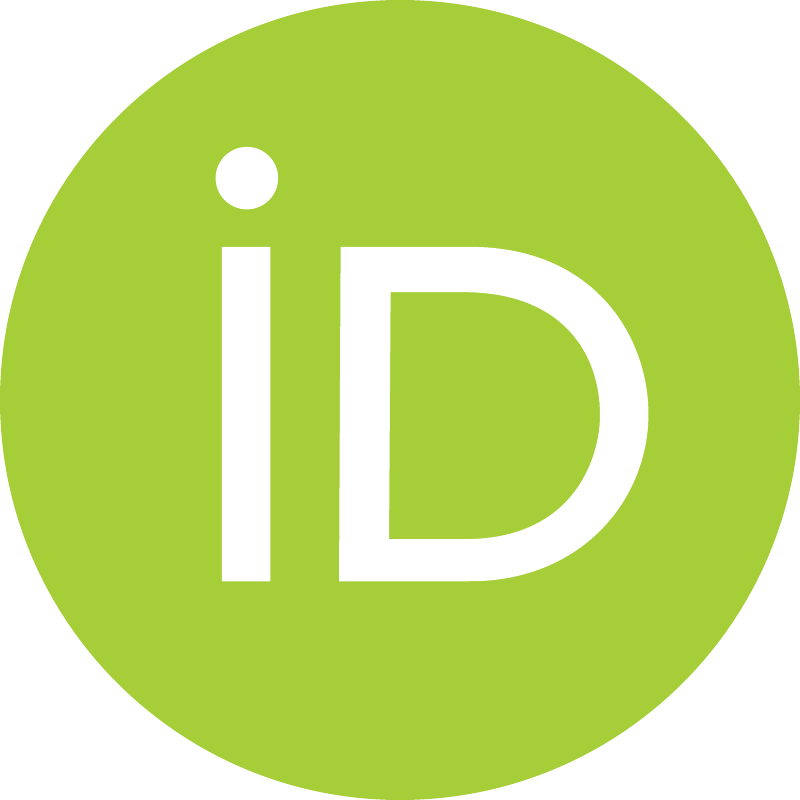}\hspace{1mm}Sergio Morell-Monzó}\thanks{Corresponding author: \texttt{sermomon@upv.es}} \\
	Underwater Acoustics Group \\
	Universitat Politècnica de València (IGIC-UPV) \\
	Carrer del Paranimf, 1 \\
	46730 Gandia, Spain \\
	\And
	\href{https://orcid.org/0000-0001-5546-3748}{\includegraphics[height=1.6em]{orcid.pdf}\hspace{1mm}Dídac Diego-Tortosa} \\
	Institut de Ciències del Mar \\
	CSIC (ICM-CSIC) \\
	Passeig Marítim de la Barceloneta 37-49 \\
	08003 Barcelona, Spain \\
	\AND
	\href{https://orcid.org/0000-0001-9010-6287}{\includegraphics[height=1.6em]{orcid.pdf}\hspace{1mm}Isabel Pérez-Arjona} \\
	Underwater Acoustics Group \\
	Universitat Politècnica de València (IGIC-UPV) \\
	Carrer del Paranimf, 1 \\
	46730 Gandia, Spain \\
	\And
	\href{https://orcid.org/0000-0001-8882-866X}{\includegraphics[height=1.6em]{orcid.pdf}\hspace{1mm}Víctor Espinosa} \\
	Underwater Acoustics Group \\
	Universitat Politècnica de València (IGIC-UPV) \\
	Carrer del Paranimf, 1 \\
	46730 Gandia, Spain \\
}

\begin{document}
\maketitle

\begin{center}
\fbox{\parbox{0.92\textwidth}{\small
This is the author version of the article published in \textit{Applied Ocean Research}, Volume 176 (2026), 105281, after peer review. DOI: \href{https://doi.org/10.1016/j.apor.2026.105281}{10.1016/j.apor.2026.105281}. Received 20 May 2026; revised 22 September 2026; accepted 23 September 2026. The content of this version corresponds to the published article; the typeset version of record is available from the publisher under the CC BY-NC 4.0 licence (\url{http://creativecommons.org/licenses/by-nc/4.0/}). Please cite the published version.}}
\end{center}

\begin{abstract}
Distributed Acoustic Sensing (DAS) enables continuous monitoring of dynamic strain along tens of kilometres of optical fibre, generating massive datasets whose interpretation and automated analysis remain challenging. DAS data can be understood as a time-distance image array. However, the absence of standardised representation methods limits both the physical interpretability and the development of automated image-based analysis pipelines. This work introduces a systematic framework for visualisation and feature extraction of DAS data based on a multispectral signal representation. The approach decomposes strain-rate measurements into predefined frequency bands and produces band-limited energy images that describe the spatial and temporal distribution of acoustic energy across distinct spectral regimes. The framework is evaluated using DAS recordings containing fin whale and blue whale vocalisations. Three experiments are conducted to assess the approach: enhanced visualisation of bioacoustic signals, unsupervised clustering of acoustic patterns, and event detection using a convolutional neural network. The visualisation experiment shows that the multispectral representation allows for the simultaneous visualisation of sets of events that cannot be visualised using the strain amplitude representation. The unsupervised clustering experiment confirms that the multispectral representation yields class-discriminative feature spaces without requiring labelled data, enabling the automatic grouping of acoustically distinct event types. Finally, the use of multispectral composites as input of a ResNet-18 achieves an accuracy of 95.9\% in whale vocalisation detection, demonstrating that the proposed representation provides an effective feature space for automated analysis. This approach provides a generalisable bridge between raw sensor data and standard computer vision pipelines, with implications for large-scale automated monitoring using fibre-optic sensor networks.
\end{abstract}


\section{Introduction}
\label{sec:intro}

Distributed Acoustic Sensing (DAS) technology has emerged over the past decade as a revolutionary tool for continuous monitoring of geophysical, environmental, and human activities along extensive fibre-optic infrastructures \citep{shang2022}. Unlike conventional sensors, which provide point measurements, DAS enables the recording of strain-rate variations continuously along tens to hundreds of kilometres of fibre with metric-scale spatial resolution and high sampling rates \citep{he2021}. The technique operates by launching coherent laser pulses into a standard optical fibre and measuring phase changes in the Rayleigh backscattered signal caused by dynamic strain perturbations. Through this approach, the fibre behaves as a dense set of spatially distributed sensing channels, allowing the wavefield to be monitored simultaneously in space and time \citep{hartog2017}.

A major operational constraint of DAS technology is the volume of data generated during acquisition. For instance, a system comprising approximately 2500 sensing channels sampled at 250~Hz can produce around 200~MB every 10~s \citep{idrissi2025}, equivalent to 1.2~GB/min. For this reason, many deployments operate below the maximum instrumental sampling rate and need near-real-time analysis algorithms to identify and save time windows containing the raw data for a more detailed post-analysis. As noted by \citet{shang2022}, automated event detection systems for DAS are not yet fully mature for many real-world applications requiring robust real-time performance. Therefore, efficient extraction of informative acoustic features is crucial for long-term monitoring, requiring compact signal descriptors for event detection and environmental characterisation \citep{diegotortosa2025}.

Unlike conventional sensors, whose outputs often have a direct and calibrated physical interpretation, DAS sensitivity depends strongly on fibre coating properties, cable design, and deployment conditions, such as whether the cable is buried, suspended, trenched, or resting on the seafloor. These factors determine the coupling efficiency between the medium and the fibre. This efficiency is not usually homogeneous along the entire length of the cable. Consequently, the detection and interpretation of acoustic events rely heavily on the signal processing and visualisation workflow used to represent the data.

To identify different types of events, such as the various vocalisations produced by baleen whales, different features are taken into account, including rhythmic intervals or call intervals, intensity or reception levels and shapes in the time-distance representation \citep{bouffaut2022}. However, the decisive characteristic for classifying baleen whales is the spectral content of the signal \citep{mcdonald2023, sirovic2017}.

Although DAS data allow for channel-by-channel analysis, their greatest potential lies in the combination of all channels into a time-distance representation of the strain field along the cable. Acoustic events are typically identified through the analysis of time-distance signatures. Band-pass filters, normalisation, dynamic compression, $f$--$k$ masking, and visualisation techniques can substantially modify the spatio-temporal representation. Therefore, any modification to these processing choices propagates directly into the spatio-temporal shape of the representation, altering the detectability and interpretability of physical events. Thus, events with different spectral content may take on an almost identical form in the time-distance matrix, making it difficult to distinguish between them. This phenomenon is observed in some vocalisations of baleen whales, where classification depends on both spectral content and spatio-temporal morphology \citep{mcdonald2023, sirovic2017}.

These observations highlight the need for visualisation strategies capable of simultaneously conveying both spatial and spectral information. In this sense, a conceptual parallel can be drawn with multispectral satellite remote sensing \citep{prasad2011}, where the combination of spectral bands has enabled the interpretation of complex imagery and the extraction of meaningful biophysical information \citep{zhong2018}. Analogously, DAS measurements can be interpreted as distributed observations of a dynamic field whose physical content is encoded in its frequency structure. Decomposing the DAS signal into physically meaningful frequency bands yields a structured multiband representation, where each band highlights events distinguishable by their frequency content and facilitates the identification of energy regimes linked to oceanographic, biological, and anthropogenic processes.

Existing approaches to DAS feature extraction and visualisation, such as band-pass filtering, are typically applied on a case-by-case basis to isolate a specific type of event or frequency band, and do not provide a systematic framework for jointly exploiting the spectral content across multiple bands simultaneously. As a result, the spectral information that could help identify and discriminate physical events is used only partially, event by event, rather than being combined into a compact and structured representation. The central hypothesis of this work is that a multispectral representation allows events with distinct frequency characteristics to be studied jointly within a single, unified representation, rather than requiring a separate processing pipeline for each event type or frequency range.

The work introduces a systematic approach for feature extraction and visualisation of DAS data grounded in a multispectral representation. The proposed taxonomy uses spectral decomposition into frequency bands to extract physical information from different acoustic events. The viability of this approach is demonstrated through three complementary experiments: (\textit{i}) enhanced visualisation of acoustic phenomena using false-colour multispectral compositions, (\textit{ii}) feature extraction and unsupervised clustering, (\textit{iii}) event detection using a simple deep learning algorithm. These experiments are conducted to detect and distinguish fin whale (\textit{Balaenoptera physalus}) and blue whale (\textit{Balaenoptera musculus}) vocalisations as a representative case study, demonstrating how the proposed framework captures biologically meaningful spectral structure and supports automatic discrimination of marine acoustic events.

\section{Multispectral representation of DAS}
\label{sec:methods}

\subsection{Spectral decomposition}
\label{sec:spectral}

Many physical phenomena observed in DAS exhibit characteristic frequency signatures. Therefore, spectral analysis constitutes a primary tool for feature extraction and analysis. Inspired by multispectral imaging approaches, the DAS data are decomposed into several frequency bands, enabling the construction of physically interpretable spectral layers.

Let $x(s,t)$ denote the DAS signal, where $s$ and $t$ represent the distance along the fibre (indexed by channel number) and time, respectively. This representation is commonly known as a \textit{waterfall}. For each spatial channel $s_i$, the temporal signal can be transformed into the frequency domain using the Fourier transform:
\begin{equation}
  X(s_i,\, f) = \int x(s_i, t)\, e^{-j2\pi ft}\, dt
  \label{eq:fourier}
\end{equation}

Eq.~(\ref{eq:fourier}) formalises the frequency content of the DAS signal and motivates its decomposition into frequency bands. Band-limited components could be obtained by masking $X(s_i, f)$ directly in the frequency domain and applying the inverse Fourier transform. Thus, the signal is decomposed into $K$ predefined frequency bands $[f_{k1},\, f_{k2},\, \ldots,\, f_{ki}]$, each designed to capture specific events. In practice, band decomposition is performed directly in the time domain using band-pass filters, applied independently to each channel. This prevents the $O(N \log N)$ computation cost and edge artefacts of an explicit Fourier-domain implementation. Since a single-pass filter introduces frequency-dependent phase distortion, each filter is applied in a forward--backward, zero-phase configuration with well-posed initial conditions \citep{gustafsson1996}. Suitable filter include Chebyshev, Bessel and Butterworth filters.
\begin{equation}
  x_k(s,t) = \mathcal{B}_k\{x(s,t)\}
  \label{eq:bandpass}
\end{equation}
where $\mathcal{B}_k$ denotes a band-pass filter for band $k$. The energy within each band is estimated as the instantaneous power of the filtered signal:
\begin{equation}
  E_k(s,t) = \left| x_k(s,t) \right|^{2}
  \label{eq:energy}
\end{equation}

Each $E_k(s,t)$ forms a two-dimensional array (distance and time). These spectral band images capture how acoustic energy is distributed across frequencies, space, and time and they can be stacked along the frequency axis.

\subsection{Visualisation pipeline}
\label{sec:pipeline}

\begin{figure}[t]
  \centering
  \includegraphics[width=\textwidth]{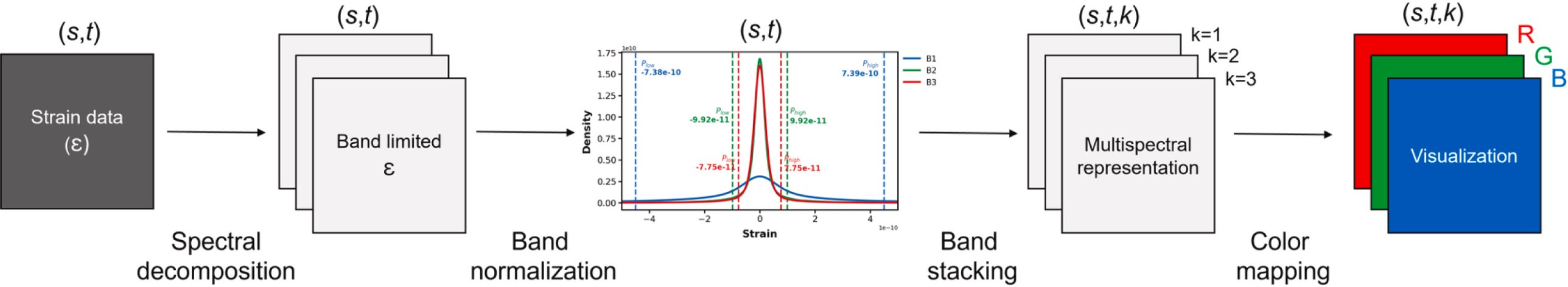}
  \caption{Scheme of the visualisation pipeline to produce RGB compositions.}
  \label{fig:fig1}
\end{figure}

The multispectral representation described above provides a structured set of band-limited energy images. In order to transform these spectral layers into interpretable visualisations, a systematic visualisation pipeline is defined. This pipeline consists of four main stages: (1) spectral decomposition, (2) normalisation, (3) band stacking and (4) colour mapping (see Fig.~\ref{fig:fig1}).

To perform multispectral visualisation, the first step is to perform the spectral decomposition as described in Section~\ref{sec:spectral}. Each channel applies a 5th order Butterworth filter as the bandpass filter. Each $E_k(s,t)$ spectral band forms a two-dimensional array representing the distribution of acoustic energy within a specific frequency interval.

Because different frequency bands may have substantially different amplitude distributions, normalisation is required prior to visualisation. Without normalisation, high-energy bands would dominate the dynamic range, masking weaker structures in other bands. Therefore, independent normalisation is applied to each band. Let $\hat{E}_k(s,t)$ denote the normalised band $k$, obtained independently for each band using robust percentile-based absolute clipping, instead of minimum-maximum normalisation. Let $v_k$ denote the $p$th percentile of $|E_k(s,t)|$, used as the reference amplitude for band $k$. The normalised band is obtained by:
\begin{equation}
  \hat{E}_k(s,t) = \operatorname{clip}\!\left( \frac{|E_k(s,t)|}{v_k},\ 0,\ 1 \right), \qquad v_k = \operatorname{percentile}_p\!\left( |E_k(s,t)| \right),
  \label{eq:norm}
\end{equation}
with $p = 95$ used throughout this work. This percentile-based clipping provides robustness to outliers compared to a simple min--max rescaling, at the cost of saturating the top $(100-p)\%$ of amplitude values to 1. This formulation implicitly assumes that $E_k(s,t)$ is approximately zero-centred and symmetrically distributed, which holds for DAS strain data band-pass filtered around a central frequency. Under this assumption, the single-sided percentile $v_k$ is equivalent to two symmetric bounds $\pm v_k$ on the signed distribution. For signals that are not zero-centred or exhibit asymmetric amplitude distributions, this formulation could be generalised using two independent percentiles, $P_\mathrm{low}$ and $P_\mathrm{high}$, of the signed (non-absolute) distribution, with normalisation performed as $(E_k(s,t) - P_\mathrm{low})/(P_\mathrm{high} - P_\mathrm{low})$ followed by clipping to $[0,1]$ in Eq.~(\ref{eq:norm}).

The set of band energy images is then stacked along a third dimension to form a multichannel spectral cube $\hat{E}(s,t,k)$. This stacked representation preserves the spectral separation of energy and constitutes the basis for both visualisation and feature extraction.

Finally, each normalised band $\hat{E}_k(s,t)$ is rescaled into a fixed interval; floating-point representations typically employ a normalised range from zero to one, whereas 8-bit image encoding utilises an intensity range from zero to 255. Then, three normalised bands are assigned to the Red (R), Green (G), and Blue (B) colour channels. The correct selection of bands determines the physical interpretation of the composite image for each type of event, and spectrally distinct phenomena appear with different colours depending on their frequency dominance. Multiple RGB composites can be generated by varying band combinations, enabling targeted visualisation of specific acoustic regimes.

\subsection{Feature extraction}
\label{sec:features}

Spectral Decomposition (SD) is not merely a visualisation tool that enhances interpretability, but a feature extraction method in its own right. Instead of treating $x(s,t)$ as a single scalar quantity at each distance--time coordinate, multispectral decomposition transforms every sample into a vector of band-derived attributes. For each position $(s,t)$, the set of band energies $\{E_1(s,t), E_2(s,t), \ldots, E_K(s,t)\}$ forms a multiband descriptor. Consequently, each pixel of the original waterfall representation is mapped into a feature space whose dimensions correspond to physically defined frequency intervals and their relationships.

SD generates a structured embedding of the DAS field in which heterogeneous acoustic events become distinguishable in the feature space. The discriminative power does not arise from a specific classifier, but from the physically informed organisation of the spectral energy itself.

\subsection{Experimental setup}
\label{sec:setup}

\subsubsection{Data}
\label{sec:data}

For the three experiments, data from the benchmark dataset \textit{RAPID: Distributed Acoustic Sensing on the OOI's Regional Cabled Array} \citep{wilcock2023data} were used. This public dataset contains recordings from one of the first DAS campaigns for marine bioacoustic monitoring, conducted in November 2021 offshore of central Oregon (USA), in the northeastern Pacific Ocean. During this deployment, vocalisations from fin whales and blue whales were successfully recorded.

The experiment was conducted on two submarine cables of the Ocean Observatories Initiative (OOI) Regional Cabled Array (RCA). The DAS measurements acquired using the OptaSense interrogator were recorded under three different configurations: C1 (sampling frequency of 200~Hz and gauge length of 50~m), C2 (sampling frequency of 500~Hz and gauge length of 30~m), and C3 (sampling frequency of 1000~Hz and gauge length of 30~m), all with a channel spacing of 2~m.

\subsubsection{Experiment 1: Visualisation}
\label{sec:exp1}

The objective of this experiment is to assess the capability of the proposed multispectral framework for enhancing the visualisation of marine bioacoustic events. RGB composites were generated from the spectral band energy images $E_k(s,t)$ following the pipeline described in Section~\ref{sec:pipeline} and compared against conventional band-limited strain waterfall representations. Traditional visualisations were obtained by filtering the strain signal $x(s,t)$ within a selected frequency band and displaying the resulting amplitude image. In contrast, the multispectral approach assigns three normalised band energies to the RGB channels, enabling frequency-dependent chromatic encoding.

Two use cases were evaluated. First, bands were selected to enhance the visualisation of fin whale 20~Hz pulses against background noise: R: 16--28~Hz, centred on the fin whale fundamental frequency, G: 30--40~Hz and B: 40--60~Hz, adjacent bands used to capture background and anthropogenic noise for contrast. Second, bands were selected to discriminate between type-A and type-B vocalisation patterns of fin and blue whales, centring each band on the fundamental or harmonic frequencies reported in the literature for these call types \citep{wilcock2023jasa} (e.g., R: 20--28~Hz, G: 16--20~Hz, B: 3--15~Hz for fin whale type-A/type-B calls; R: 12--18~Hz, G: 26--32~Hz, B: 40--48~Hz for the blue whale type-B harmonics). This experiment focuses on the qualitative assessment of visual separability and interpretability provided by the multispectral composition.

\subsubsection{Experiment 2: Feature extraction and unsupervised clustering}
\label{sec:exp2}

To illustrate the descriptive capacity of the multispectral representation, a simple unsupervised segmentation experiment was conducted using the $k$-means algorithm on multispectral composites built from three bands (16--28~Hz, 30--40~Hz, 40--60~Hz), matching those used in Experiment~1. The experiment consists of segmenting the time--distance matrix, classifying each of its pixels into different groups. The quality of the segmentation depends directly on the structure of the multispectral features. If the band decomposition captures physically meaningful contrasts between frequency regimes, different acoustic phenomena naturally form coherent clusters without requiring supervision.

\subsubsection{Experiment 3: Event detection}
\label{sec:exp3}

The third experiment evaluates the suitability of the proposed multispectral representation for supervised event detection. A Convolutional Neural Network (CNN) based on the ResNet-18 architecture \citep{he2015} --- implemented using the TorchVision library for PyTorch \citep{torchvision2016} --- was trained as a classifier to determine the presence or absence of whale vocalisations within a given time--distance window. The input to the network consists of three-band multispectral composites derived from normalised energy images $\hat{E}_k(s,t)$. These bands are stacked to form a three-channel image, analogous to an RGB representation, enabling direct use of standard CNN architectures without structural modification.

The C1 data subset used for this experiment contains data recorded between November 3, 2021 at 22:08:02~UTC and November 5, 2021 at 14:06:02~UTC. Data were registered with the OptaSense interrogator at a sampling frequency of 200~Hz, with a spatial sampling interval of 2~m and a gauge length of 50~m. The data is stored in one-minute intervals. Before further processing, original arrays were truncated to the first 16\,300 channels (0--33.283~km), and only these parts were considered in the analysis. Each file was converted to an RGB image using the same band assignment as in Experiment~2: 16--28~Hz, 30--40~Hz, and 40--60~Hz. Images were then rescaled to minimise memory usage during training: images whose longest axis exceeded 1024 pixels were resized to this threshold, preserving the original aspect ratio through Lanczos resampling. This subset produced a total of 2394 images, rescaled to a size of 1024 $\times$ 681 pixels.

To assess the stability of the results and avoid bias from a single data partition, the model was evaluated using repeated stratified cross-validation, with 5 folds and 6 repetitions (30 independent runs in total). In each run, the dataset was partitioned into training (60\%), validation (20\%), and test (20\%) subsets, all stratified by class to preserve the original class proportions, resulting in approximately 1436 training images, 479 validation images, and 479 test images per run. The validation subset was used for early stopping (patience of 8 epochs without improvement in validation accuracy, up to a maximum of 25 epochs).

The distribution demonstrates a pronounced imbalance favouring the non-vocalisation class, with approximately 78\% compared to 22\%. This pattern is consistent across all subsets due to class-stratified sampling. Preserving the natural class proportions in each partition, rather than artificially balancing them, ensures that the reported metrics reflect the conditions the classifier would face in an operational monitoring setting. This imbalance reflects typical conditions in passive acoustic and DAS monitoring. Detectable events occur only during short periods, when a whale passes through the cable's receptive field and vocalises. Consequently, most time windows contain no vocalisations.

The ResNet-18 model was trained using the AdamW optimiser, with a learning rate of $1 \times 10^{-4}$ and a weight decay of $1 \times 10^{-4}$. A step learning rate scheduler was applied, reducing the learning rate by a factor of 0.5 every 7 epochs. The batch size was set to 4. The categorical cross-entropy loss was used as the optimisation objective. The ResNet-18 backbone was initialised with weights pretrained on ImageNet, and the final fully connected layer was replaced and trained from scratch to match the two-class output. All model parameters remained trainable during fine-tuning. Each of the 30 runs used a distinct random seed (derived from a base seed of 42, incremented by the run index), applied to both the training/validation split and model initialisation, ensuring that individual runs were reproducible while differing from one another. Data augmentation, comprising random horizontal and vertical flips, was applied to the training subset only; the validation and test subsets were not augmented. The performance of the models was evaluated using the overall accuracy, precision, recall, and F1-score. This experiment assesses whether the spectral decomposition provides a discriminative feature space from which a standard deep learning model can effectively learn biologically meaningful patterns.

\section{Results}
\label{sec:results}

\subsection{Visualisation}
\label{sec:res_vis}

\begin{figure}[t]
  \centering
  \includegraphics[width=\textwidth]{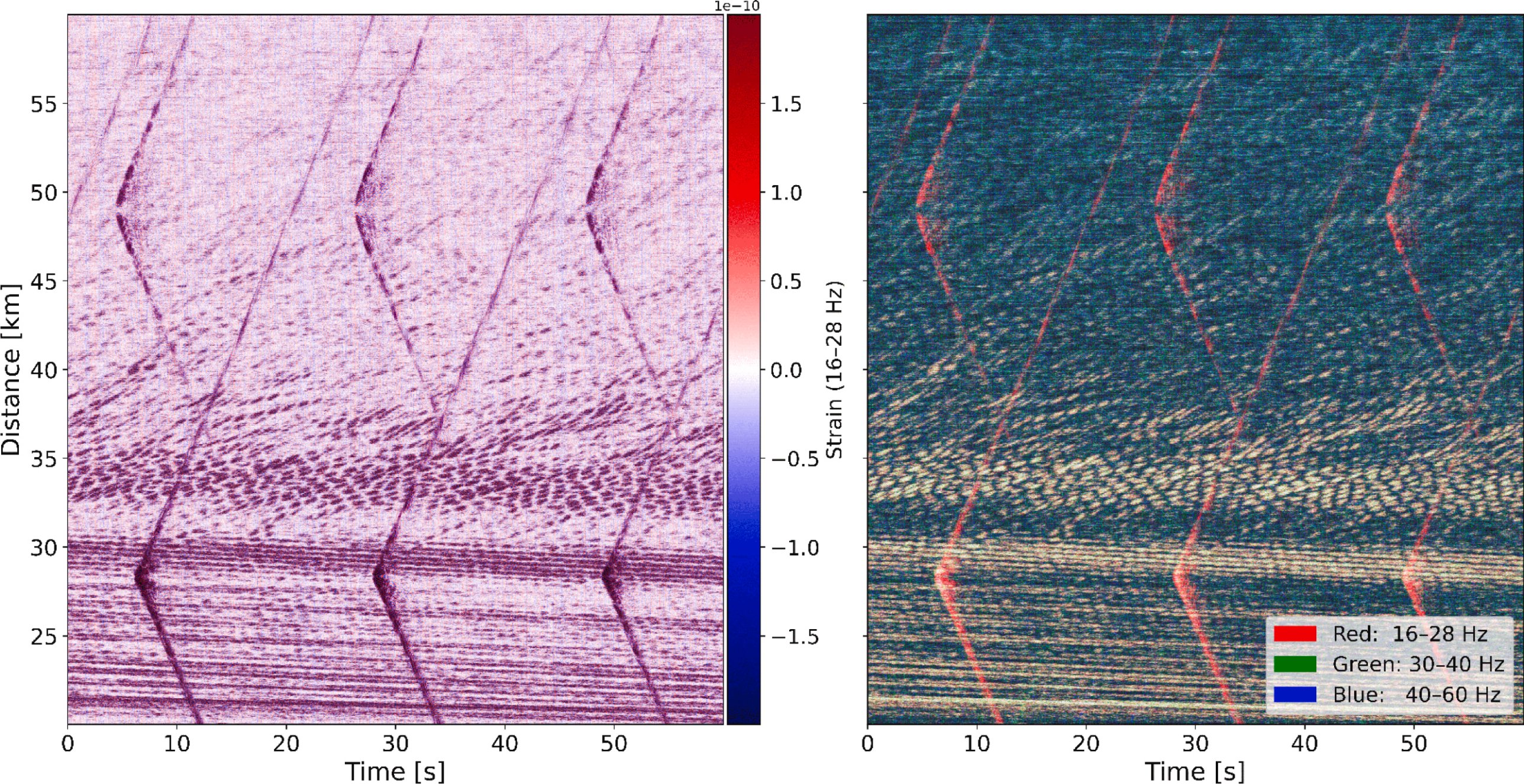}
  \caption{Comparison between traditional single-band visualisation and multispectral representation for the same data segment acquired on November 4, 2021, at 19:30:02~UTC. Left: single-band strain waterfall filtered in the 16--28~Hz band. Right: multispectral representation combining three frequency bands as an RGB composite (R: 16--28~Hz, G: 30--40~Hz, B: 40--60~Hz), where colour intensity reflects the normalised energy limited to each band. The dynamic range of both panels was clipped at the 95th percentile of the respective amplitude/energy distribution to enhance visual contrast.}
  \label{fig:fig2}
\end{figure}

Fig.~\ref{fig:fig2} compares the conventional band-limited strain visualisation with the proposed multispectral representation for a 60~s data segment containing three fin whale vocalisations. In the left panel, the DAS data are filtered within the 16--28~Hz band and displayed using a traditional single-band amplitude representation. In the right panel, the multispectral representation is shown by assigning three frequency bands to the RGB channels: R (16--28~Hz), G (30--40~Hz), and B (40--60~Hz). In both representations, the characteristic V-shaped patterns from a powerful emitter are clearly visible. These structures arise from the differential arrival times of the acoustic pulse along the fibre, reflecting the spatial propagation of the signal.

However, in the traditional band-limited visualisation, fin whale vocalisations and background noise exhibit similar intensity levels and are therefore displayed with comparable colour tones, limiting visual separability. In contrast, the multispectral representation enhances spectral discrimination. Fin whale 20~Hz pulses, with observed values typically ranging from approximately 16 to 28~Hz \citep{kingnolan2024}, are primarily mapped to the R channel and thus appear distinctly red. Other background acoustic components associated with maritime traffic and fibre optic noise (mostly between 0 and 40~km), with relatively higher contributions in the G and B bands, are rendered in greyish-green tones. This chromatic separation illustrates how multispectral decomposition facilitates visual differentiation between biologically generated signals and ambient or anthropogenic noise, even when amplitude levels are comparable.

\begin{figure}[t]
  \centering
  \includegraphics[width=\textwidth]{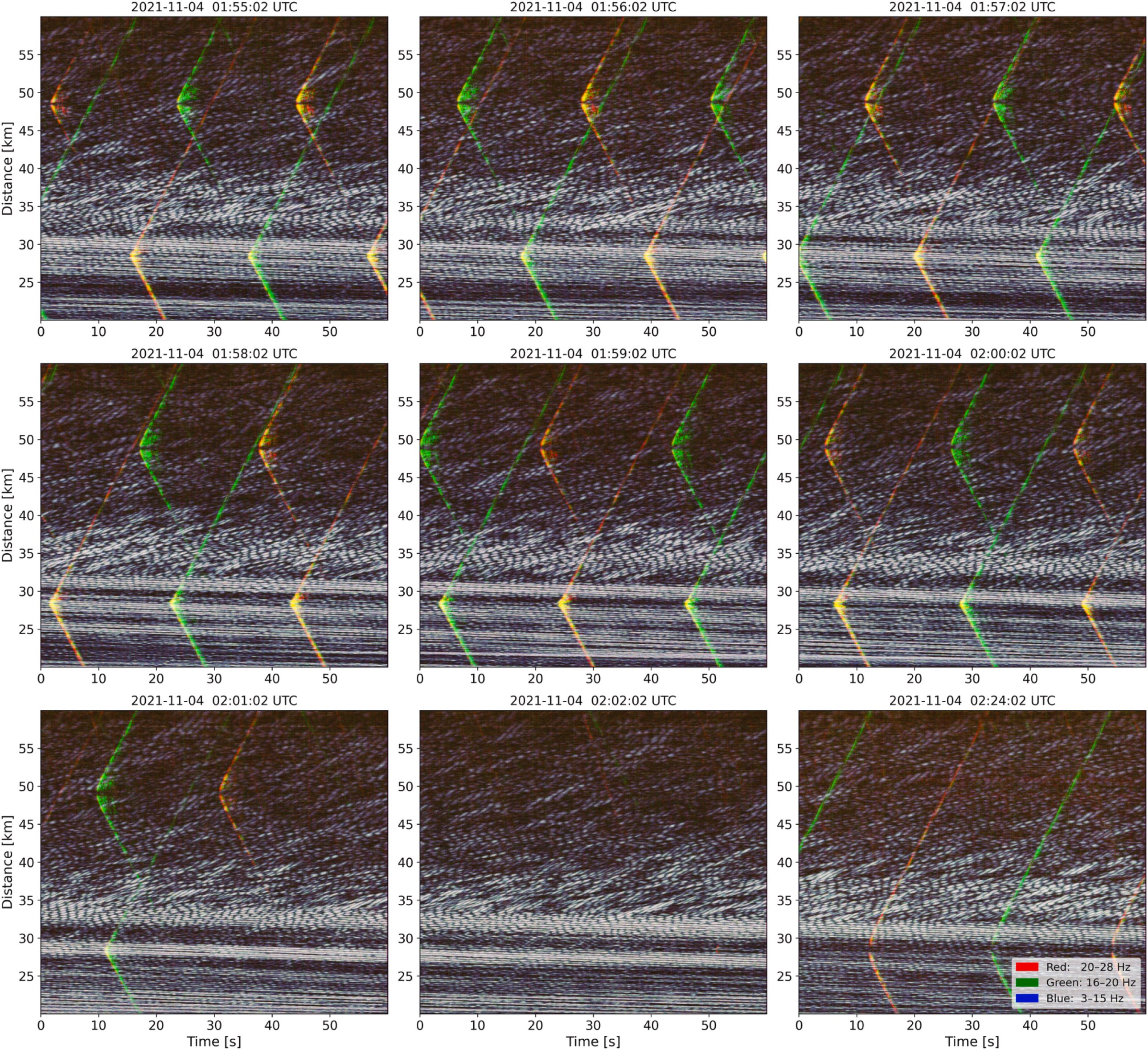}
  \caption{Multispectral representation of a collection of fin whale 20~Hz pulses across nine successive one-minute time windows (November 4, 2021, 01:55--02:02~UTC, plus a subsequent window at 02:24~UTC for comparison purposes), illustrating the temporal evolution and repetition of the sounds. Type-A (orange) and type-B (green) vocalisations are visible as distinct coloured traces; the orange colouration of some type-A calls results from partial spectral overlap between the red and green bands. The blue channel (3--15~Hz) captures low-frequency background noise, providing visual context for the calls. The greyish/whitish tones visible throughout the panel correspond to broadband noise present across all three bands, including noise inherent to the cable.}
  \label{fig:fig3}
\end{figure}

Fig.~\ref{fig:fig3} shows a collection of fin whale 20~Hz pulses recorded during the same experiment. In this case, the multispectral representation was configured to enhance intra-species spectral differences by assigning the 16--20~Hz band to the G channel and the 20--28~Hz band to the R channel. This configuration enables clear discrimination between the two primary vocalisation types in the 20~Hz notes emitted by the fin whale, commonly referred to as type-B and type-A calls \citep{sciacca2015, weirathmueller2017}. Type-B vocalisations, with dominant energy in the 16--20~Hz range, appear predominantly green, whereas type-A vocalisations, whose spectral energy peaks between 20 and 28~Hz, are rendered in orange tones due to the combined contribution of the R and G channels.

Both vocalisation types are clearly distinguishable from the acoustic background, which exhibits comparatively lower energy within these narrow low-frequency bands. The resulting chromatic contrast highlights subtle spectral differences that would be difficult to identify in a single-band representation. This example demonstrates that the proposed multispectral framework not only enhances event visibility but also provides a physically interpretable basis for intra-class discrimination, supporting its potential use as a feature representation for automated classification tasks.

There is a clear alternation between type-A and type-B fin whale vocalisations in Fig.~\ref{fig:fig3}. This structured pattern is immediately apparent in the multispectral representation, which highlights not only spectral differences but also biologically meaningful temporal organisation.

\begin{figure}[t]
  \centering
  \includegraphics[width=\textwidth]{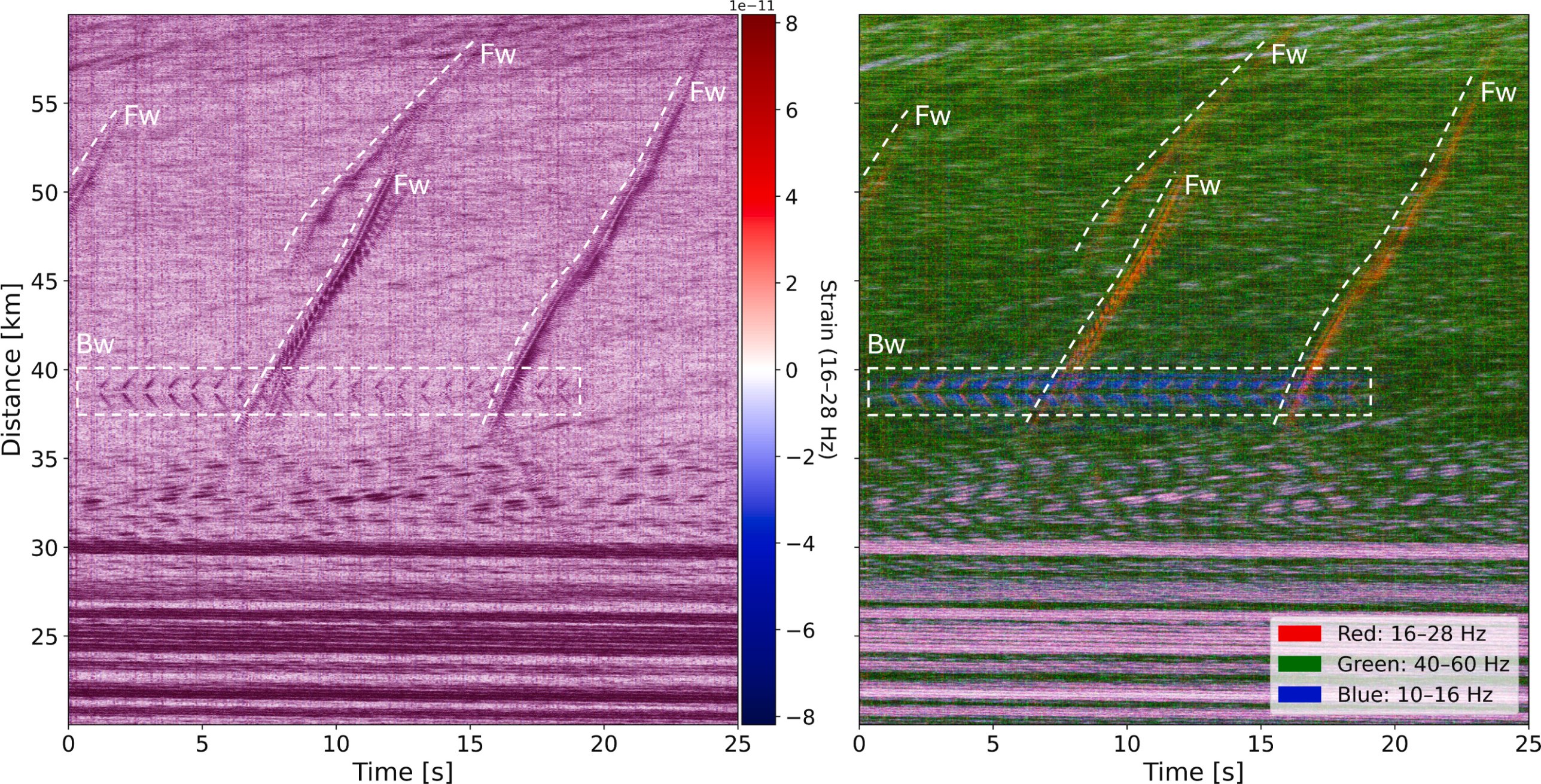}
  \caption{Recording from November 2, 2021, at 10:36:14~UTC containing a type-A blue whale (Bw) sound, with the nearest source along the cable located approximately 38~km away, overlaid by fin whale (Fw) pulses at 20~Hz. Left: single-band strain waterfall filtered in the 16--28~Hz band. Right: Multispectral representation combining three frequency bands as an RGB composite (R: 16--28~Hz, G: 40--60~Hz, B: 10--16~Hz). The blue whale sound is shown in blue, while the fin whale sounds are shown in red. The dynamic ranges of both graphs were adjusted to the 95th percentile to allow for a fair comparison.}
  \label{fig:fig4}
\end{figure}

Multispectral compositions with appropriately selected bands also enable the visualisation of vocalisations from different species simultaneously. Fig.~\ref{fig:fig4} shows a recording containing fin whale 20~Hz notes alongside a type-A blue whale vocalisation of approximately twenty seconds in duration. The type-A blue whale call exhibits a pulsed structure, appearing as a regular train of short, repeated pulses confined to a narrow distance range around 38--40~km, in contrast to the fin whale notes, which appear as isolated arrivals whose apex marks the closest approach of the animal to the cable. In the single-band representation (left), filtered in the 16--28~Hz range, both call types are visible on the same colour scale: although this band is centred on the fin whale notes, the broadband nature of the blue whale pulses leaves a clear imprint within it, so that the two sources can only be told apart by their morphology. In the multispectral representation (right), the two frequency ranges are decomposed into separate bands and mapped onto different colour channels: fin whale notes in red (16--28~Hz) and blue whale pulses in blue (10--16~Hz), which allows both to be identified simultaneously and unambiguously, even where they intersect in time and distance. The green channel (40--60~Hz) contains no biological signal of interest in this case and captures background and vessel noise, providing spatial context for the calls. The pinkish tones observed in the first kilometres of the array arise from the additive combination of comparable energy in the red and blue channels with a weaker green contribution, indicating that the acoustic energy in that section is concentrated at low frequencies across a broad range rather than confined to a single band.

\begin{figure}[t]
  \centering
  \includegraphics[width=\textwidth]{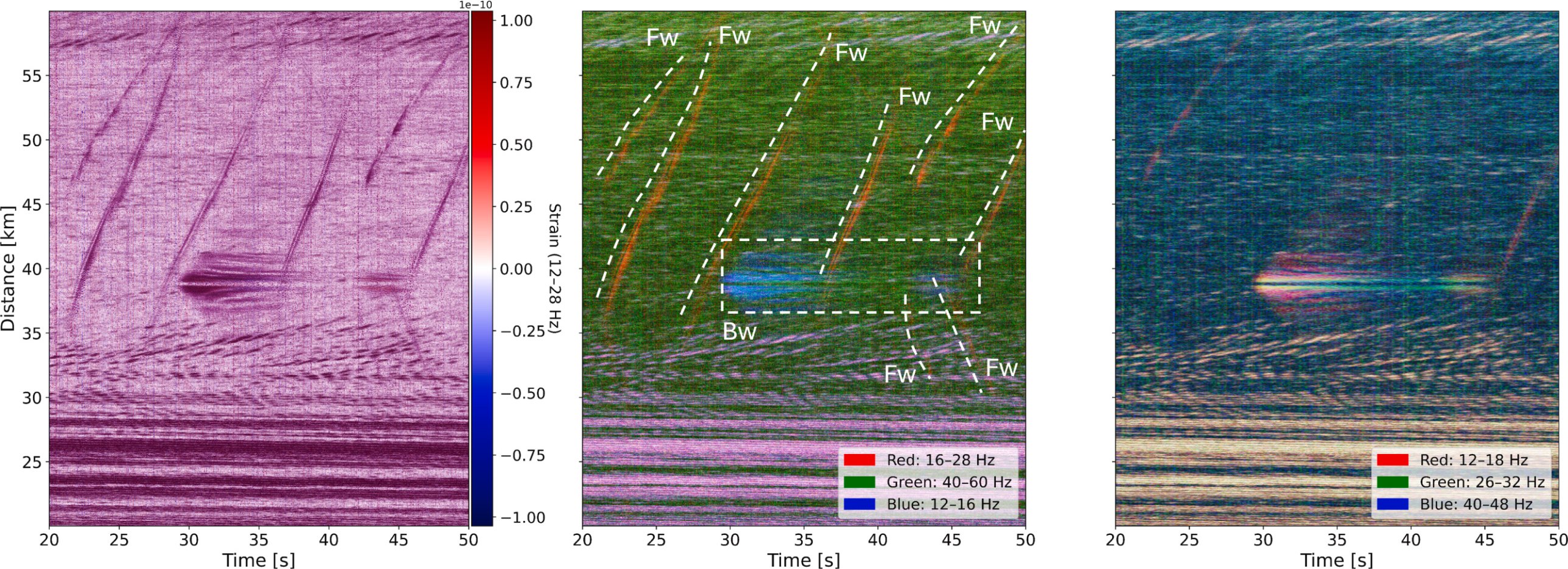}
  \caption{Recording from November 2, 2021, at 10:33:14~UTC, containing a type-B (Bw) blue whale sound, with the nearest source along the cable located approximately 38~km away, overlaid by fin whale (Fw) sounds at 20~Hz. Left: single-band deformation cascade filtered in the 12--28~Hz band. Centre: multispectral representation combining three frequency bands into an RGB composite (R: 16--28~Hz, G: 40--60~Hz, B: 12--16~Hz), where Bw and Fw sounds are shown in blue and red, respectively. Right: Multispectral representation combining three frequency bands into an RGB composite (R: 12--18~Hz, G: 26--32~Hz, B: 40--48~Hz), where each harmonic of the Bw sound has been assigned to a colour channel to form a rainbow pattern. The dynamic ranges of all graphs were adjusted to the 95th percentile to allow for a fair comparison.}
  \label{fig:fig5}
\end{figure}

Fig.~\ref{fig:fig5} shows a type-B blue whale vocalisation co-occurring with several fin whale 20~Hz notes \citep{wilcock2023jasa}. In the single-band representation (left), filtered in the 12--28~Hz range so as to encompass both the blue whale and the fin whale frequency ranges, the blue whale vocalisation is clearly visible around 38--40~km, but it is rendered on the same colour scale as the surrounding fin whale notes and can therefore only be told apart by its morphology and duration. The two multispectral compositions illustrate how the choice of bands determines what the representation makes explicit. In the central panel, the bands are selected to separate species: the blue whale fundamental (12--16~Hz) is mapped to the blue channel and the fin whale notes (16--28~Hz) to the red channel, so that both sources are simultaneously identifiable even where their arrivals intersect. In the right panel, the bands are instead centred on the first three harmonics of the type-B call (12--18~Hz, 26--32~Hz and 40--48~Hz), which makes its internal harmonic structure explicit. Notably, type-B blue whale vocalisations exhibit a distinctive rainbow-like chromatic pattern due to the presence of multiple harmonics, with characteristic frequencies centred near 15~Hz, 30~Hz, and 45~Hz \citep{weirathmueller2017}. It is worth noting that blue whale vocalisations are scarce in this dataset, and that a single-band representation conveys only their morphology; the multispectral representation adds a spectral dimension to that visual inspection, so that the harmonic structure of the call becomes a chromatic signature that can be recognised at a glance and unambiguously distinguished from co-occurring vocalisations of other species.

\subsection{Feature extraction}
\label{sec:res_feat}

\begin{figure}[t]
  \centering
  \includegraphics[width=\textwidth]{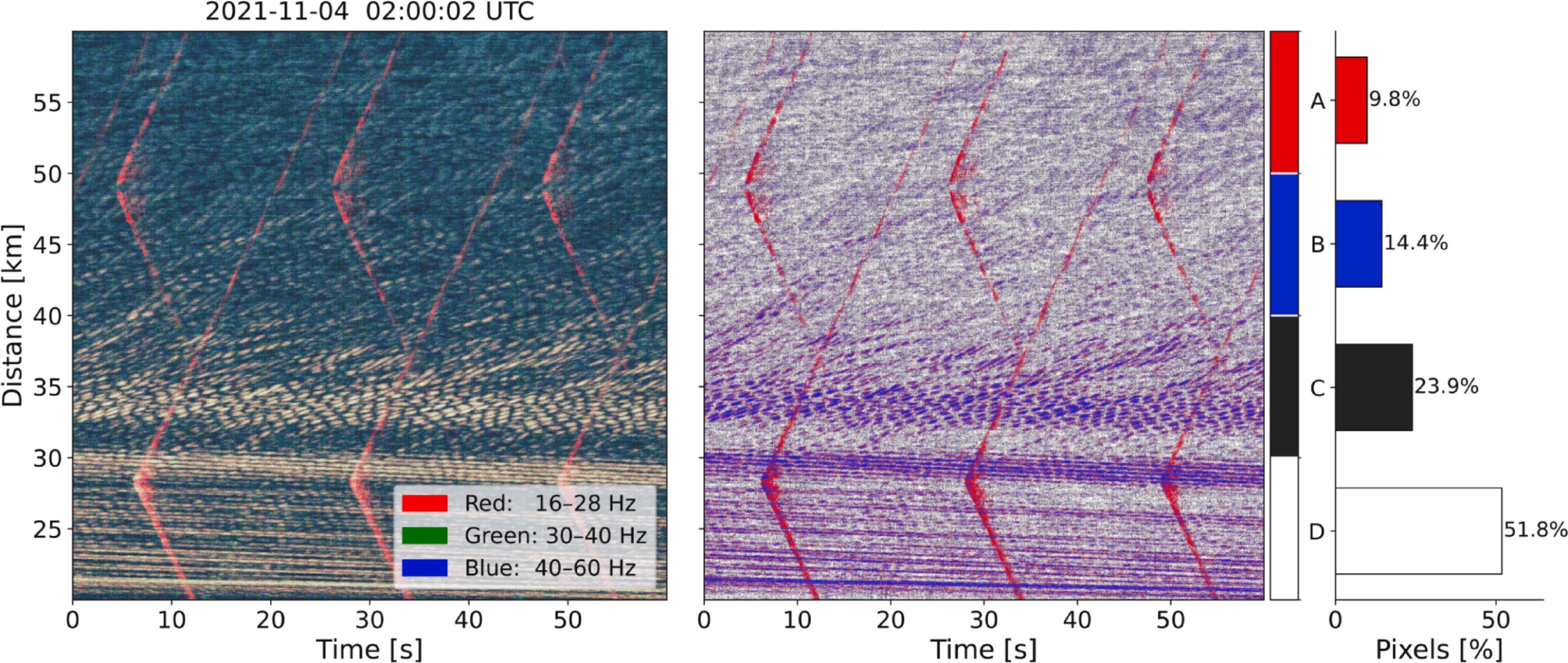}
  \caption{Results of unsupervised clustering with $k$-means into four groups. The multispectral representation is shown on the left and the segmented image on the right. The bar charts show the percentage of pixels in each cluster (A, red: fin whale vocalisations; B, blue: structured broadband noise; C, black, and D, white: diffuse background).}
  \label{fig:fig6}
\end{figure}

Fig.~\ref{fig:fig6} presents the results of unsupervised clustering using $k$-means into four groups, based on the multispectral representation derived from three spectral bands (B1: 16--28~Hz, B2: 30--40~Hz, B3: 40--60~Hz). The left panel shows the original multispectral image, while the right panel shows the segmented output. The clustering was performed on a pixel-wise basis, where each pixel is represented by its multispectral feature vector.

The resulting segmentation reveals coherent spatial--temporal regions that correspond to distinct acoustic patterns, and the characteristic V-shaped propagation patterns are preserved. Fin whale 20~Hz vocalisations, background noise, and other acoustic structures are grouped into separate clusters, demonstrating that the multispectral representation provides sufficient spectral contrast to enable unsupervised separation without prior labelling.

Regarding Fig.~\ref{fig:fig6}, cluster A primarily captures pixels associated with the fin whale notes, accounting for 9.8\% of the total pixels. Cluster B aggregates the structured broadband noise present across all three bands, including the horizontal striping associated with the cable itself, representing 14.4\% of the pixels. Clusters C and D contain the remaining diffuse background and pixels near class boundaries, corresponding to 23.9\% and 51.8\% of the pixels, respectively. The segmentation, however, is far from perfect due to the use of a simple $k$-means algorithm. It also operates on a pixel-wise basis, with no spatial regularisation term. Each pixel is therefore assigned independently of its neighbours. The speckle noise in the image propagates directly into the segmentation.

These results indicate that the multispectral composition effectively captures the internal structure of the DAS data and serves as a robust feature extraction method. Furthermore, the inclusion of additional spectral bands would expand the feature space, potentially enhancing the discriminative power of the clustering by improving the separability of overlapping acoustic events.

\subsection{Event detection}
\label{sec:res_det}

The ResNet-18 model underwent training for a maximum of 25 epochs per run, employing early stopping with a patience threshold of 8 epochs without improvement in validation accuracy. Across the 30 cross-validation runs, training concluded between epochs 8 and 24, with an average termination at approximately 15 epochs. The multispectral model demonstrated a mean accuracy of 95.9\% $\pm$ 1.0\% (95\% CI [95.5\%, 96.3\%]). The mean precision and recall were 95.9\% $\pm$ 0.9\% and 95.9\% $\pm$ 1.0\%, respectively. Table~\ref{tab:metrics} provides a summary of the overall and class-wise performance metrics, along with their variability across runs. Additionally, the pooled confusion matrix is displayed in Table~\ref{tab:confusion}.

\begin{table}[t]
  \centering
  \caption{Mean and standard deviation of the classification metrics across 5-fold stratified cross-validation repeated six times (30 runs in total). Class 0 indicates the absence of whales, while Class 1 denotes their presence.}
  \label{tab:metrics}
  \begin{tabular}{lccc}
    \toprule
    Metric & Mean & Std & 95\% CI \\
    \midrule
    Accuracy             & 0.959 & 0.010 & [0.955, 0.963] \\
    Precision (weighted) & 0.959 & 0.009 & [0.956, 0.963] \\
    Recall (weighted)    & 0.959 & 0.010 & [0.955, 0.963] \\
    F1-score (weighted)  & 0.959 & 0.010 & [0.955, 0.963] \\
    \midrule
    Precision (class 0)  & 0.971 & 0.011 & [0.967, 0.976] \\
    Recall (class 0)     & 0.976 & 0.011 & [0.972, 0.981] \\
    F1-score (class 0)   & 0.974 & 0.006 & [0.972, 0.976] \\
    Precision (class 1)  & 0.917 & 0.034 & [0.904, 0.930] \\
    Recall (class 1)     & 0.898 & 0.043 & [0.882, 0.914] \\
    F1-score (class 1)   & 0.906 & 0.023 & [0.898, 0.915] \\
    \bottomrule
  \end{tabular}
\end{table}

\begin{table}[t]
  \centering
  \caption{Pooled confusion matrix obtained by aggregating predictions from the six iterations of the 5-fold cross-validation. Each iteration represents a complete pass over the dataset.}
  \label{tab:confusion}
  \begin{tabular}{llcc}
    \toprule
     & & \multicolumn{2}{c}{Predicted} \\
    \cmidrule(lr){3-4}
     & & 0 (absent) & 1 (present) \\
    \midrule
    \multirow{2}{*}{True} & 0 (absent)  & 1822 & 44  \\
                          & 1 (present) & 54   & 474 \\
    \bottomrule
  \end{tabular}
\end{table}

The results show a strong and stable classification performance for both classes across the 30 runs. Class 0 (whale-absent files) exhibits consistently high recall (0.976 $\pm$ 0.011) and a low standard deviation, indicating that the model rarely misclassifies non-vocalisation files regardless of the specific data partition. For Class 1 (whale-present files), precision remains high (0.917 $\pm$ 0.034) but recall is both lower and considerably more variable across runs (0.898 $\pm$ 0.043), suggesting that some vocalisation events are consistently missed and that this behaviour is sensitive to which samples fall into training versus test in each partition. These false negatives generally correspond to particularly challenging cases where vocalisations occupy a very small portion of the scene or appear very faint. In such situations, the signals can be difficult to visually identify even for human annotators. In addition, weak traces may be further attenuated by pre-processing steps such as image resampling and conversion to 8-bit intensity values, which reduce the dynamic range of the representation and can obscure subtle acoustic patterns. This degradation could be mitigated by using higher-bit images, improved resampling algorithms, larger image sizes, or tiled inference schemes that process smaller spatial--temporal windows.

\section{Discussion}
\label{sec:discussion}

The results demonstrate that the proposed multispectral representation provides a structured and physically interpretable framework for analysing DAS data. By decomposing the signal into frequency bands and representing the corresponding energy distribution, the acoustic field is transformed from a single-amplitude representation into a multiband description in which different physical processes can be examined separately. This transformation facilitates the visual and automatic identification of acoustic phenomena whose spectral signatures would otherwise be difficult to distinguish using conventional single-band visualisations.

The multispectral representation can jointly render acoustic phenomena that share similar time--distance morphology but differ in frequency content, without requiring separate single-band inspections. For automated analysis, this same property translates into a discriminative advantage: CNNs are known to exploit colour-channel information as a discriminative feature from the earliest layers \citep{krizhevsky2012, zeiler2014}, so the spectral information encoded as colour in the multispectral composites is, in principle, directly accessible to a CNN classifier in the same way it is accessible to a human interpreter, without requiring the network to separately learn to disentangle overlapping frequency content from a single-band amplitude image.

Beyond visualisation, multispectral decomposition provides a natural framework for feature extraction. Mapping each pixel into a vector of band-limited energies is what allows the same representation to support the three complementary analyses carried out in this work (i.e. visual inspection, unsupervised clustering, and supervised classification) without requiring a separate feature-engineering step for each. As demonstrated by the clustering experiment (Fig.~\ref{fig:fig6}), because $k$-means has no learned parameters, the coherent groups it recovers must originate in the physically informed organisation of the spectral energy itself, rather than in any classifier-specific inductive bias. This corroborates, with an unsupervised method, the same discriminative capacity that the ResNet-18 experiment later confirms under supervision, and complements other existing image-based detection techniques \citep{feliutena2022}.

The supervised detection experiment further confirms the suitability of the multispectral representation for automated analysis. Because the multispectral composites can be encoded as conventional 8-bit RGB images, they are directly compatible with the input format expected by a wide range of computer vision architectures, including object detection frameworks such as YOLO, Mask R-CNN or RT-DETR. This compatibility allows such models to be initialised from weights pretrained on large-scale datasets such as ImageNet or COCO, as exploited in Experiment~3, rather than requiring architecture-specific input layers or training from scratch. The use of three bands was a design decision to allow transfer learning, but could be extended to an arbitrary number of frequency bands, enabling multispectral or hyperspectral input representations. Multispectral representations have recently been used as input data for CNNs in related contexts, as demonstrated by \citet{zhang2026}, \citet{pierau2025}, and could complement existing detection methods such as that of \citet{truong2025}; however, a direct quantitative comparison against these approaches was not performed in this work, as it would require access to comparable datasets and annotation protocols, and is left as a direction for future work.

The proposed framework is particularly relevant for marine bioacoustics applications. Large cetacean vocalisations typically occur at low frequencies and can propagate over long distances in the ocean, allowing them to couple efficiently with submarine fibre-optic cables \citep{bouffaut2023}. At the same time, the marine acoustic environment contains a mixture of biological signals, oceanic background noise, and anthropogenic interference. The suitability of the proposed approach for this domain is further reinforced by the fact that mysticete vocalisations are primarily distinguished by their frequency content, which the multispectral representation encodes directly and explicitly. Other classes of transient sources monitored with DAS, such as seismic or anthropogenic events, may also exhibit characteristic frequency signatures, suggesting that the approach could extend beyond marine bioacoustics; however, this has not been evaluated in the present work and should be regarded as a hypothesis for future investigation.

It is necessary to acknowledge some limitations of the current approach. First, the method relies on prior expert knowledge to select the relevant frequency bands for constructing the multispectral composites. Second, all experiments were performed on a single dataset with a limited number of vocalisation types, so the approach should be evaluated on other vocalisation types and acoustic events. Third, as discussed in Section~\ref{sec:res_det}, the conversion of the multispectral composites to 8-bit RGB images and the associated resampling reduce the dynamic range of the representation, which can attenuate weak or subtle spectral traces and contribute to missed detections. Finally, the computational cost of the proposed method, both in processing time and in GPU memory requirements, increases with the number of frequency bands considered, which may become a practical constraint for high-resolution multiband representations or for real-time processing pipelines.

A natural extension of this approach is to replace the reduced bank of Butterworth filters with a decomposition based on the short-time Fourier transform (STFT), in which each channel is characterised by its power spectral density across a set of contiguous frequency bins, yielding a three-dimensional time--distance--frequency representation. Both alternatives lead to multispectral or hyperspectral representations, but with different trade-offs. The STFT provides a uniform and exhaustive sampling of the spectral axis, and its reduction in the number of time samples also acts as a useful compression of the data; its main drawback is that it imposes a single time--frequency resolution trade-off for the whole spectrum, determined by the window length, with long windows favouring the separation of narrow bands and harmonic structures and short windows better preserving the pulsed nature of certain vocalisations. The Butterworth filter bank, by contrast, retains the original temporal resolution and allows bands of non-uniform width to be tailored to the ranges of interest of each problem; in exchange, its computational cost grows linearly with the number of bands and, since each filter output is an oscillating signal, it requires an additional envelope estimation step that implicitly reintroduces an analysis time scale. The choice between the two strategies thus depends on the spectral and temporal characteristics of the events to be represented, and constitutes in itself a design parameter that dedicated architectures could determine automatically.

Future work should explore learnable band-selection strategies, in which a dedicated neural architecture identifies the relevant frequency bands internally as part of the training process, rather than relying on manually predefined bands. Since deep learning models generally perform best when allowed to learn their own feature representations, such an architecture could potentially uncover frequency patterns of interest that are not apparent from prior domain knowledge. Combined with the compatibility of the multispectral representation with standard computer-vision architectures and pretrained weights, this direction could also support the development of scalable, near-real-time monitoring pipelines for large-scale DAS deployments.

Overall, the results suggest that multispectral decomposition provides a principled basis for structuring DAS multichannel observations in a way that is both physically interpretable and compatible with modern computer vision approaches. The proposed methodology represents a step forward in the development of automated analysis systems capable of leveraging the growing volume of DAS observations.

\section{Conclusions}
\label{sec:conclusions}

This work introduces the multispectral representation framework for DAS data, grounded in the spectral decomposition of strain-rate measurements into physically defined frequency bands. By computing band-limited energy images and assigning them to RGB colour channels, the proposed approach transforms the conventional single-amplitude waterfall representation into a structured, multiband description of the acoustic field. This description is both interpretable and compatible with modern image-based analysis pipelines.

Three complementary experiments conducted on DAS recordings of marine mammal vocalisations demonstrate the versatility of the framework. First, the visualisation experiments show that multispectral compositions substantially enhance the perceptual separability of bioacoustic events from background noise. They also enable the simultaneous chromatic discrimination of spectrally distinct vocalisation types, including type-A and type-B calls of fin whale (\textit{Balaenoptera physalus}) and blue whale (\textit{Balaenoptera musculus}). These types appear with a similar morphology in a conventional single-band representation. Second, the unsupervised clustering experiment demonstrates that the multispectral feature vectors derived from spectral decomposition provide sufficient discriminative structure to separate coherent acoustic patterns without any labelled data, using a simple $k$-means algorithm. Third, the supervised event detection experiment confirms that multispectral composites constitute an effective input representation for CNNs. A standard ResNet-18 classifier trained on three-band composites achieves a mean accuracy of 95.9\% $\pm$ 1.0\%.

Beyond these quantitative results, this work proposes a novel physically interpretable representation of DAS data that connects raw DAS signals with standard image-based analysis pipelines. The framework is not specific to marine bioacoustics. Its extension to other DAS monitoring scenarios involving geophysical, environmental, or anthropogenic sources remains to be evaluated.

\section*{CRediT authorship contribution statement}

\textbf{Sergio Morell-Monzó:} Writing -- original draft, Visualization, Validation, Software, Methodology, Conceptualization. \textbf{Dídac Diego-Tortosa:} Writing -- review \& editing, Visualization, Data curation. \textbf{Isabel Pérez-Arjona:} Project administration, Funding acquisition. \textbf{Víctor Espinosa:} Writing -- review \& editing, Supervision, Project administration.

\section*{Declaration of competing interest}

The authors declare that they have no known competing financial interests or personal relationships that could have appeared to influence the work reported in this paper.

\section*{Acknowledgements}

The authors thank Dr. William Wilcock for providing a high-quality public dataset \citep{wilcock2023data} and the das4whales community for developing open-source tools to analyse it \citep{bouffaut2022data}. The contribution of Dr. Morell-Monzó was supported by the Universitat Politècnica de València through the Access Contracts for Post-Doctoral Researchers (PAID-10-25): MAR.IA -- Modelos de inteligencia artificial para el análisis de datos de acústica submarina. The contribution of Dr. Diego-Tortosa was supported by the ALLIES Cofund Program which received funding from the European Union's Horizon-MSCA-2022-COFUND-01 research and innovation programme under the Marie Skłodowska-Curie grant agreement No 101126626.

\section*{Supplementary data}

Supplementary material related to this article can be found online at \url{https://doi.org/10.1016/j.apor.2026.105281}.

\section*{Data availability}

The data used are publicly available through the RAPID experiment website: \textit{Distributed Acoustic Sensing on the OOI's Regional Cabled Array} \citep{wilcock2023data}. Source code is available on GitHub: \url{https://github.com/sermomon/MsR-DAS}.

\end{document}